\documentclass[pra,twocolumn,showpacs]{revtex4}
\usepackage{graphicx,amsmath}

\begin{document}

\title{Berezinskii-Kosterlitz-Thouless transition of two-dimensional Bose gases in a synthetic magnetic field}
\author{Junjun Xu and Qiang Gu}
\affiliation{Department of Physics, University of Science and Technology Beijing, Beijing 100083, China}
\date{\today}
\begin{abstract}
We study the Berezinskii-Kosterlitz-Thouless transition of
two-dimensional Bose gases in a synthetic magnetic field using the
standard Metropolis Monte Carlo method. The system is described by
the frustrated XY model and the critical temperature is calculated
though the absence of central peak of the wave function in momentum
space, which can be directly measured by the time-of-flight
absorbing imaging in cold atoms experiments. The results of our work
show agreement with former studies on superconducting Josephson
arrays.
\end{abstract}
\maketitle

\section{introduction} 
It is well known that in two-dimensional (2D)
systems with a continuous symmetry the conventional long-range order
is prevented by thermal fluctuations at finite temperature in the
thermodynamic limit, and as a result no spontaneous continuous
symmetry breaking takes place. However, these systems can undergo a
transition through binding of vortex pairs to form a
quasi-long-range order. This is the celebrated
Berezinskii-Kosterlitz-Thouless (BKT) transition
\cite{Berezinskii,Kosterlitz}. Theoretically, the 2D XY model is the
prototype to elaborate on the BKT transition. Experimentally, the
BKT transition has been examined in various physicsl systems,
including $\rm ^4He$ films \cite{Bishop}, 2D superconductors
\cite{Beasley} and superconducting Josephson-junction arrays (JJA)
\cite{Resnick}, and experimental works agree well with theoretical
predictions.

The study of Josephson-junction arrays greatly enriched the research
of the BKT transition \cite{Resnick, Voss, Teitel, Shih, Brown}.
Once a transverse magnetic field is introduced to JJA, the BKT
transition point as a function of the field can be investigated
\cite{Resnick, Voss}. Theoretically, such a system can be described by the
frustrated 2D XY model \cite{Teitel, Shih, Brown}. Nevertheless,
these systems reveal a complex energy spectrum, known as the 
``Hofstadter butterfly" \cite{Hofstadter},
which gives rise to rich phenomena that are still of present research interest.

So far, cold atoms have been regarded as ideal test beds for
fundamental models of condensed matter physics. It is of great
interest to examine the BKT transition since the quasi-2D Bose gas
has already been produced either in a single pancake traps or at the
nodes of one-dimensional optical lattice potentials \cite{Bloch}.
Trombettoni {it et al.} has raised a proposal to investigate the
occurrence of the BKT transition in a 2D scalar Bose-Einstein condensate
\cite{Trombettoni}. The BKT transition in atomic Bose gases was first
experimentally realized in 2006 by Hadzibabic \cite{Hadzibabic}.
They added the optical lattice in the $z$ direction to construct a
quasi-two-dimensional configuration, and the transition was detected
by the dislocations of interference patterns. Later experiments
identify this transition by the proliferation of vortex pairs
\cite{Schweikhard} and a bimodal density distribution or
coherent length \cite{Clade}.

The newly developed technic of synthetic magnetic field in cold
atoms use the light-modulated excitation between nearby neighbor
sites to construct a "magnetic field" like effective hamiltonian in
a 2D optical lattice \cite{Lin}. This technic opens the door for
cold atoms to simulate some more profound models, e.g. the lattice
gauge theory. Thus it stimulates us to re-examine the behavior
of BKT transition in the cold atomic Bose gases subject to the synthetic 
magnetic field. Comparing with the former 2D experimental systems 
concerning on the superfluid density or sheet resistance to detect 
the superfluid transition, cold atoms can be manipulated more easier 
and precisely and thus it benefits for further investigation of 
the BKT transition.

We start with the Bose-Hubbard model which describes a 2D Bose gas 
in uniform magnetic field in a square optical lattice. 
This model can be mapped to the
frustrated XY model. Then we study the BKT transition using the standard
Metropolis Monte Carlo method. This method has been used by Trombettoni
\emph{et al.} to investigate the BKT transition in a atomic Bose gases
in a 2D optical lattice without magnetic field \cite{Trombettoni}. 
The transition point is determined through detecting the central peak 
of the expansion condensate after time-of-flight (TOF) expansion. 

\section {the model}
Consider a 2D Bose gas immersed into a uniform magnetic field in a square 
optical lattice. The Hamiltonian of this system can be described by the
frustrated Bose-Hubbard model:
\begin{equation}
  \hat{H}=-K\sum_{\left\langle i,j\right\rangle}\left(\hat{\psi}_i^\dag\hat{\psi}_je^{iA_{ij}}+h.c.\right)+\frac{U}{2}\hat{N}_i\left(\hat{N}_i-1\right),
\label{eq:e1}
\end{equation}
where the first term on the right side is the hopping energy with
the summation over the nearest neighbor sides and $K$ is the hopping
matrix element. $\hat{\psi}_i$($\hat{\psi}_i^{\dag}$) is the field
operator of bosons for annihilate(create) a boson at site $i$.
$A_{ij}$ is a bond operator describing the magnetic field and around
every plaquette we have:
\begin{equation}
  A_{ab}+A_{bc}+A_{cd}+A_{de}=2\pi f.
\end{equation}
Here $f=\frac{2e}{hc}H_0a^2=\frac{H_0}{\phi_0}$ is the uniform
frustration with $\phi_0=hc/2e$ the flux quantum and we have chosen
unit lattice constant $a=1$. The second term of Eq.~(\ref{eq:e1})
refers to the on-site repulsive(attractive) interaction, according
to the interacting strength $U>0(U<0)$, and  $\hat{N}_i$ is the
particle number operator on site $i$.

The system under consideration is a uniform system at sufficient low
temperature, such that the system is density coherent with
fluctuating phases. The field operator of bosons can be written as
$\psi_i=\sqrt{N_0}e^{i\theta_i}$, where $N_0$ is the average
particle number of each site and $\theta_i$ is the corresponding
phase on site $i$. The mechanism of BKT transition, the pairing of
vortices, is due to topological long-range correlation, thus we can
safely ignore the on-site interaction term at sufficient low
temperature. Upon these, the Eq.~(\ref{eq:e1}) can be mapped to the
frustrated XY model:
\begin{equation}
  \hat{H}=-J\sum_{\langle i,j\rangle}\cos\left(\theta_i-\theta_j+A_{ij}\right).
  \label{eq:e2}
\end{equation}
Here $J=2KN_0$. This model was first used by Titel \emph{et al.} to
describe the superconducting Josephson arrays in transverse magnetic
field \cite{Teitel}.

We note here that the frustrated XY model is $U(1)$ gauge symmetry
breaking in the superfluid state as the hopping term breaks the
conservation of "charge". It is to say that in a conventional system
described by the frustrated XY model physical quantities can be
gauge dependent, however, all observable quantities should be gauge
invariant. However this is not the case in the cold atoms experiment
that the observable physical quantities, e.g., the momentum
distribution of the wave function are gauge dependent. There is no
paradox and on the other hand they are compatible in the case that
the imaging of density of the expanding condensates in cold atoms
experiment are in fact the canonical momentum of the original model
\cite{Moller}. This just reflects nothing but the exitance of vector
gauge potential.

In the case of synthetic magnetic field we choose the Landau gauge, which is :
\begin{equation}
  \textbf{A}=\left(0,2\pi fx\right).
\end{equation}
Obviously the hamiltonian in Eq.~(\ref{eq:e2}) is periodic in $f$
with the period 1, thus we only need to study the properties of the
system in the interval $f\in[0,1]$. Here $f=0$ corresponds to the
unfrustrated case and $f=\frac{1}{2}$ is the fully frustrated
condition.

The wave function of the system in the lattice space $\psi_{i}$ can
be calculated using the Monte Carlo method. At low temperature the
system undergo a superfluid transition and there will be a peak in
the central of the wave function in the momentum space, which is
fourier transformed as:
\begin{equation}
  \tilde{\psi}_{\textbf{k}}=\frac{1}{N_s}\sum_{j}\psi_{j}e^{-i\textbf{k}\cdot\textbf{r}_j},
\end{equation}
where $N_s$ is the number of sites of the square lattice. The
central peak of the momentum space describes the coherence of the
phase of the condensates. It is analogous to the magnetization of a
spin system, i.e.,
\begin{equation}
  M=\langle\tilde{\psi_0}\rangle.
\end{equation}

In cold atoms experiments, the observation of momentum distribution
can be detected by the sudden release of the optical lattice. The
absorption imaging is taken after a TOF period $t$. The
density profile of the image can be written as \cite{Bloch}:
\begin{equation}
    n(\textbf{x})=(M/\hbar t)^3 |\tilde{\omega}(\textbf{k})|^2 G(\textbf{k}).
\end{equation}
Here momentum $\textbf{k}$ is related to position $\textbf{x}$ by $\textbf{k}=M\textbf{x}/\hbar t$ under the assumption of ballistic expansion. $\tilde{\omega}(\textbf{k})$ is the Fourier transform of the Wannier function and $G(\textbf{k})$ is the momentum space density matrix and is defined by:
\begin{equation}
    G(\textbf{k})=\frac{1}{N_s} \sum_{i,j}e^{i\textbf{k}\cdot(\textbf{r}_i-\textbf{r}_j)} \langle\hat{\psi}_i^\dag\hat{\psi}_j\rangle.
\end{equation}

For density coherent states at low temperature, the system forms a
quasi-condensate where on each site of the lattice there is a small
condensate and we can have
$\langle\hat{\psi}_i^\dag\hat{\psi}_j\rangle \approx
\psi_i^\ast\psi_j$. Thus the momentum space density matrix
$G(\textbf{k})$ can be written as:
\begin{eqnarray}
    G(\textbf{k})&&\approx\frac{1}{N_s} \sum_{i,j}e^{i\textbf{k}\cdot(\textbf{r}_i-\textbf{r}_j)} \psi_i^\ast\psi_j \nonumber\\
    &&=N_s\cdot |\tilde{\psi}_{\textbf{k}}|^2.
\end{eqnarray}

In this case, the momentum space wave function is related to the
density matrix and then can be observed experimentally using the
TOF imaging. In the following we will do the Monte Carlo
simulation and investigate the transition behavior under different
frustration $f$, which can be directed realized in cold atoms
experiments.

\section {Results and discussions} 
We now use the standard Metropolis 
Monte Carlo method with periodical boundary conditions to simulate
the frustrated XY model. It has been used to investigate the BKT
transition in Josephson arrays by Titel. This method is proved to
always get believable results. The lattice sites are chosen as:
$L\times L=40\times 40$. For each temperature and frustration of the
system we use $10^7$ Monte Carlo steps.

\begin{figure}[tb]
  \includegraphics[width=0.45\textwidth]{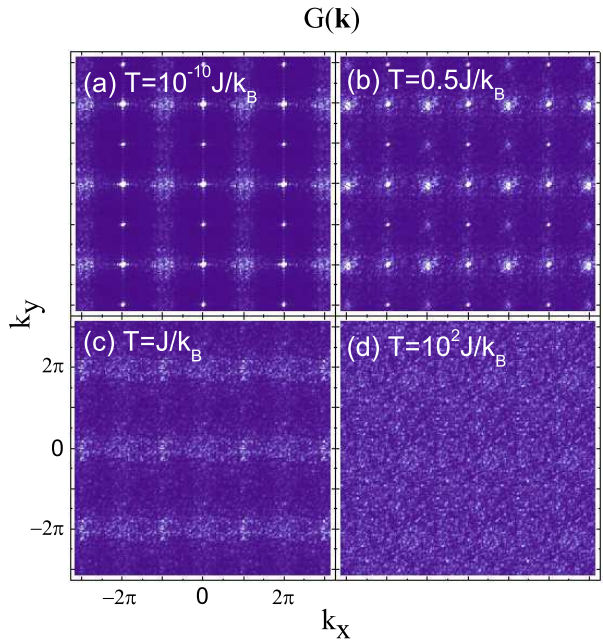}
  \caption{An illustration of expansion image of the system for different temperature $T$ at frustration $f=1/2$. The BKT transition is taken place at about $T=0.5J/k_B$ as in fig.b.}
  \label{fig:fig1}
\end{figure}

Fig.~\ref{fig:fig1} illustrates the evolution of expansion image
from phase coherent ground state to the non-coherent state. At
sufficient low temperature the system is in a superfluid state and
the density profile reveals regular sharp peaks at the momentum
space lattices. That is due to the pairing of vortices. As the
temperature increasing the peaks begin to decay and at the
transition point at about $Tc\approx0.5J/k_B$ the sharp peaks go to
zero. That's the signature of BKT transition of 2D Bose gases. We
can see that above the critical temperature there is disordered in
the density profile of the momentum space and the system loses its
coherent phase.

\begin{figure}[tb]
  \includegraphics[width=0.48\textwidth]{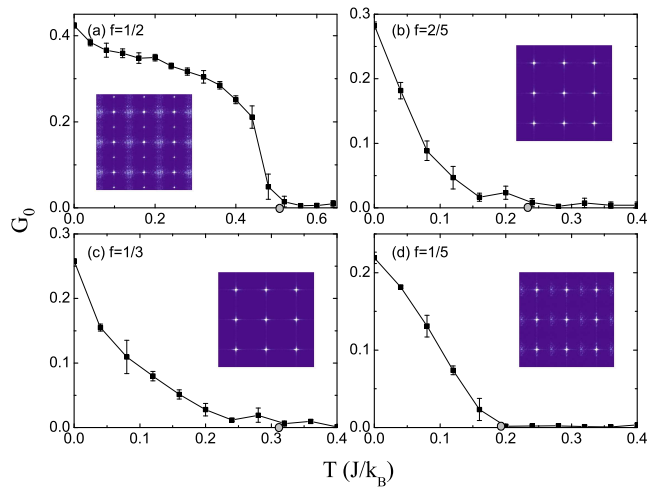}
  \caption{The central peak $G_0$($G(k_x=0,k_y=0)$) of expansion image as a function of temperature $T$ for four different fractional frustration $f$. The insert is the corresponding image close to the ground state. The square points are the numerical results with the error bar the standard deviation and the circle in each figure guides the critical transition temperature.}
  \label{fig:fig2}
\end{figure}

In Fig.~\ref{fig:fig2} we show the central peak
$G_0$($G(k_x=0,k_y=0)$) of expansion image as a function of
temperature $T$ for four different fractional frustration $f$. At
$T=0$ the peaks have maximal values where the system is in a
superfluid state with paired vortices and nonzero $M$. As the
temperature is growing, the peaks are reduced and the vortices begin
to unpair. The central peak drops to zero at the critical point and
the transition temperature is guided by the circle. The inserts
illustrate the ground state image for different frustration. For the
fully frustrated case in Fig.~\ref{fig:fig2}a we can see the curve
reveals different tendency as in Fig.~\ref{fig:fig2}b and
Fig.~\ref{fig:fig2}c. The expansion images in the latter case are
the same as the non-frustration system. We note that ground state
has $1/f$ degenerate states while this can not be directly revealed
from the expansion image. However, they can be recognized using
printing phase technology \cite{Moller}.

\begin{figure}[tbh]
  \includegraphics[width=0.45\textwidth]{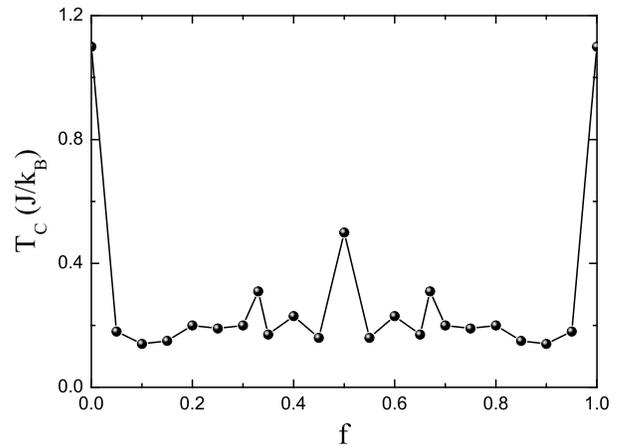}
  \caption{Transition temperature $T_c$ of 2D BKT transition for different fractional frustration $f$. The critical point is corresponding to the absence of the central as illustrated in Fig.~\ref{fig:fig2}.}
  \label{fig:fig3}
\end{figure}

We show the critical temperature $T_c$ at different fractional
frustration $f$ in Fig.~\ref{fig:fig3}. It reveals a unanalytical
behavior of the transition temperature. We can see that there are
peaks in some fractional magnetic field. The largest peak is the
fully frustrated case in $f=\frac{1}{2}$ while the second largest is
in $f=\frac{1}{3}$, which agrees with former experimental and
theoretical work on superconducting Josephson arrays \cite{Teitel}.
We note that the structure of the diagram depends on the
geometry of the lattice. For a triangular structure the experimental
and theoretical works reveal a second peak for the transition
temperature in $f=\frac{1}{4}$ instead \cite{Shih, Brown}.

\section {Conclusions} 
In conclusion, we have studied the BKT transition
of 2D Bose gases in the synthetic magnetic field using the standard
Metropolis Monte Carlo method. The critical transition temperature is
decided by the absence of central peak of the density matrix in
momentum space, which can be detected in cold atoms by TOF expansion
imaging. We have got the overall phase diagram of BKT transition for
different fractional frustration. The obtained results agree with former 
experimental and theoretical work on superconducting Josephson arrays \cite{Teitel}.
Our results suggest that cold atoms can be used as an ideal system to explore
the physics of the frustrated 2D systems. 

{\it Note added:} After this paper was almost finished, a 
preprint by Y. Nakano, K. Kasamatsu, and T. Matsui appeared on the arXiv.org 
\cite{Nakano}. These authors have studied the finite-temperature phase 
structures of hard-core bosons in a two-dimensional optical
lattice subject to an effective magnetic field based on the
extensive Monte Carlo simulations. 

\begin{acknowledgments}
This work was supported by the National Natural Science Foundation
of China (Grant No. 11074021), the Specialized Research Fund for the 
Doctoral Program of Higher Education (No. 20100006110021), and the Fundamental Research
Funds for the Central Universities of China. 
\end{acknowledgments}


\begin{thebibliography}{99}

\bibitem{Berezinskii} V. L. Berezinskii, Sov. Phys. JETP 34, 610 (1972);

\bibitem{Kosterlitz} J. M. Kosterlitz and D. J. Thouless, J. Phys. C {\bf 6}, 1181 (1973).

\bibitem{Bishop} D. J. Bishop and J. D. Reppy, Phys. Rev. Lett. {\bf 40}, 1727 (1978).

\bibitem{Beasley} M. R. Beasley, J. E. Mooij, and T. P. Orlando, Phys. Rev. Lett. {\bf 42}, 1165 (1979).

\bibitem{Resnick} D. J. Resnick, J. C. Garland, J. T. Boyd, S. Shoemaker, and R. S. Newrock, Phys. Rev. Lett. {\bf 47}, 1542 (1981).

\bibitem{Voss} R. F. Voss, and R. W. Webb, Phys. Rev. B {\bf 25}, 3446 (1982).

\bibitem{Teitel} S. Teitel and C. Jayaprakash, Phys. Rev. B {\bf 27}, 598 (1983); S. Teitel and C. Jayaprakash, Phys. Rev. Lett. {\bf 51}, 1999 (1983).

\bibitem{Shih} W. Y. Shih and D. Stroud, Phys. Rev. B {\bf 28}, 6575 (1983).

\bibitem{Brown} R. K. Brown and J. C. Garland, Phys. Rev. B {\bf 33}, 7827 (1986).

\bibitem{Hofstadter} D. R. Hofstadter, Phys. Rev. B {\bf 14}, 2239 (1976).

\bibitem{Bloch} I. Bloch, J. Dalibard, and W. Zwerger, Rev. Mod. Phys. {\bf 80}, 885 (2008).

\bibitem{Trombettoni} A. Trombettoni, A. Smerzi, and P. Sodano, New J. Phys. {\bf 7}, 57 (2005).

\bibitem{Hadzibabic} Z. Hadzibabic, P. Kr\"{u}ger, M. Cheneau, B. Battelier, and J. Dalibard, Nature {\bf 441}, 1118 (2006).

\bibitem{Schweikhard} V. Schweikhard, S. Tung, and E. A. Cornell, Phys. Rev. Lett. {\bf 99}, 030401 (2007).

\bibitem{Clade} P. Clad\'{e}, C. Ryu, A. Ramanathan, K. Helmerson, and W. D. Philips, Phys. Rev. Lett. {\bf 102}, 170401 (2009).

\bibitem{Lin} Y. J. Lin, R. L. Compton, K. Jim\'{e}nez-Garc\'{i}a, J. V. Porto, and I. B. Spielman, Nature {\bf 462}, 628 (2009).

\bibitem{Moller} G. M\"{o}ller and N. R. Cooper, Phys. Rev. A {\bf 82}, 063625 (2010).

\bibitem{Nakano} Y. Nakano, K. Kasamatsu, and T. Matsui, arXiv: 1112.0145 (2011).

\end{thebibliography}
\end{document}